\documentclass[twocolumn,superscriptaddress,aps]{revtex4-1}

\usepackage{color}
\usepackage{graphicx} 
\usepackage{epstopdf}
\usepackage{placeins}
\usepackage{dblfloatfix}
\usepackage{subfloat}
\usepackage{mathpazo}
\usepackage{diagbox}
\usepackage{mathtools}

\usepackage{array}

\newcolumntype{P}[1]{>{\centering\arraybackslash}p{#1}}

\newcommand {\be} {\begin{equation}} 
\newcommand {\ba}{\begin{eqnarray}} 
\newcommand {\ee} {\end{equation}} 
\newcommand{\ea} {\end{eqnarray}}

\renewcommand{\epsilon}{\varepsilon}

\begin{document}

\title{Spin Polarization of Photoelectrons in GaAs Excited by Twisted Photons}
\vspace*{15px}
\author{Maria Solyanik-Gorgone}

\affiliation{Department of Physics,
The George Washington University, Washington, DC 20052, USA}

\author{Andrei Afanasev}
\affiliation{Department of Physics,
The George Washington University, Washington, DC 20052, USA}

\date{\today}

\begin{abstract}

Inter-band photo-excitation of electron states with the twisted photons in GaAs, a direct band-gap bulk semiconductor, is considered theoretically. Assuming linearity of the quantum transition amplitudes and applying Wigner-Eckart theorem, we derive a plane-wave expansion of twisted-photon amplitudes. We also obtain relative probabilities for magnetic sub-level population of the photo-electrons in conduction band. The approach for calculating the position dependent electron polarization, resulting from photo-absorption of twisted light, is described for vertical transitions in the $\Gamma$ - point. Theoretical predictions for GaAs show modification of the magnitude and the sign of photoelectron polarization in the region near photon's phase singularity.

\end{abstract}

\maketitle

\section{Introduction}

The interest in Orbital Angular Momentum (OAM)  states of light, or twisted light, has been continuously growing since the publication by Allen et al in 1992 \cite{Allen:1992zz}. Within the last two decades considerable progress has been made towards understanding the mechanism of photon OAM transfer to the external and internal degrees of freedom in various systems, including turbid media \cite{jantzi2018enhanced, mamani2018transmission}, bulk semiconductors \cite{clayburn2013search}, and solid state heterostructures \cite{miao2016orbital}. For recent reviews see, {\it e.g.} \cite{Yao11, torres2011twisted, andrews2011structured, Padgett2015, Franke-Arnold2017, padgett2017orbital}.

In this paper we study polarization of photoelectrons caused by the quanta of OAM light in the conduction band of bulk GaAs with zincblende lattice. Due to spin-orbit effects in its band structure, GaAs is a material of choice in spin-polarized electron sources for electron accelerators, polarization electron microscopy and spintronics  \cite{HernandezGarcia:2008zz}. Bulk GaAs has a theoretical limit of 50\% polarization of photoelectrons when exposed to circularly-polarized light \cite{pierce1976photoemission}. In order to achieve higher polarization, up to the theoretical limit of 100\%, mechanical strain is applied \cite{Maruyama91}. However, strained or super-lattice photocathodes are associated with a number of technological and operational challenges and have limited quantum efficiency \cite{HernandezGarcia:2008zz}. Motivated by the need of robust and efficient polarized-electron source, the authors of Ref.\cite{clayburn2013search} tested a hypothesis that OAM of light can result in additional spin polarization of photoelectrons in GaAs. They obtained a null result: no effect of OAM on photoelectron polarization was observed.

To the best of our knowledge, there were no prior calculations of electron polarization in semiconductors excited by OAM light. Semi-classical formalism for OAM-light photo-excitations in bulk semiconductor and semiconductor heterostructures was previously developed in \cite{quinteiro2009electric, quinteiro2009theory, quinteiro2010twisted, cygorek2015insensitivity}, where electron kinetics in the conduction band has been addressed. In contrast to these previous studies, we focus on impact parameter dependence of the photo-excited electron states. In our theoretical model we make use of L\"owdin's theoretical formalism for conduction band electron wave functions \cite{voon2009kp, elder2011double, luttinger1955motion}, which allows to express Bloch states of conduction electrons in terms of two separate contributions - remote and near states. This approach enables direct coupling of photon total angular momentum to a near conduction electron state though Clebsch-Gordan coefficients. In a way this approach is similar to single ion photo-excitaions.

As for the OAM light, we express a quantum mechanical state of the twisted photon in an angular spectrum representation \cite{Afanasev2013kaa}, relating the twisted amplitudes to plane wave ones in a factorized form. This formalism has already successfully proved itself by describing transfer of OAM to the electrons bound in a single ion in a Paul trap \cite{rodrigues2016excitation, peshkov2017photoexcitation, afanasev2018experimental}. The corresponding experimental measurements \cite{schmiegelow2012light, schmiegelow2015excitation} are characterized by high, nanometer-scale, resolution of target ions. In the same way as topologically structured light reveals details of ions' structure in free space\cite{ afanasev2018experimental}, it is expected that the local band structure in solid state materials will be probed with higher resolution. Experimental studies of semiconductor fluorescence patterns, similar to such classical experiments as \cite{seymour1980time}, but triggered by the OAM light, may enable access to the information about underlying symmetries in semiconductors. This is also of high importance in polarization and perturbation spectroscopy of luminescent species.

The paper is structured as follows: in section II we will outline the earlier developed L\"owdin's formalism for the photo-excitations in bulk semiconductors, triggered by zero-OAM light. We will revisit the methods and limitations for applying the Wigner-Eckart theorem to the electronic states in valence and conduction bands in GaAs. Section III is focused on applying the plane-wave approach to photon states with OAM. In section IV we simplify our treatment by neglecting remote electron states and provide theoretical predictions of the resulting position-dependent electron polarization pattern. We draw conclusions and outline the prospectives in section V.
\vfill

\section{Photo-absorption in Bulk Semiconductors}
\subsection{Conventional Formalism: Overview}

In this section we consider the case with the direct-band type semiconductor $v-c$ transition being photo-excited by a zero-OAM laser beam. The general photo-absorption matrix element is known to be \cite{kittel1963quantum, ridley2013quantum}

\begin{equation}
M_{vc}^{\mu} = \int_V u_{k_c} (\pmb{r}) e^{-i \pmb{k_c} \cdot \pmb{r}} \underbracket{ e^{i \pmb{q} \cdot \pmb{r}} (\hat{\epsilon}_{\pmb{q} \lambda}} \cdot \pmb{p}) u_{k_v} (\pmb{r}) e^{i \pmb{k_v} \cdot \pmb{r}} d^3 r
\label{05/18/2018_2}
\end{equation}
where $u_{k} (\pmb{r})$ are the Bloch states; $\hat{\epsilon}_{\pmb{q} \lambda}$ is the photon polarization state and $\pmb{k}_c, \pmb{k}_v, \pmb{q}$ are the electron, hole and photon wavenumbers correspondingly. The term in the under-bracket corresponds to the plane wave vector field. Here one usually makes a common assumption of $\pmb{k}_v, \pmb{k}_c, \pmb{q}$ being much less than the zone boundary momentum $\pi/a_0$ ($a_0=5.65 \text{\AA}$ for GaAs).

\begin{equation}
q << \pi/a_0
\label{08/09/2018_4}
\end{equation}
Hence the transition matrix can be approximated as
\begin{equation}
M_{vc}^{\mu} = \sum_{\ell} e^{i(\pmb{k}_v - \pmb{k}_c + \pmb{q}) \cdot \pmb{R}_{\ell}} \int_{\text{cell}} u_{k_c} (\pmb{r}) (\hat{\epsilon}_{\pmb{q} \lambda} \cdot \pmb{p}) u_{k_v} (\pmb{r}) d^3r
\end{equation}
or in terms of the momentum matrix element,
\begin{equation}
M_{vc}^{\mu} = \sum_{\ell} e^{i(\pmb{k}_v - \pmb{k}_c + \pmb{q}) \cdot \pmb{R}_{\ell}} \; (\epsilon_{\pmb{q} \lambda})_{\mu} p_{vc}^{\mu}
\label{05/18/2018_1}
\end{equation}
with $\pmb{R}_{\ell}$ being the vector pointing to the center of the $\ell$-th cell in the crystal. The momentum matrix is typically extracted from the perturbation theory together with the dispersion relations for the band structure. Forbidden transitions proportional to the $\int d^2r \;u_c u_v$ have been neglected assuming the orthonormality of the Bloch states.

\subsection{Hamiltonian Matrix Element}

In our formalism we will assume that the electronic state in a bulk semiconductor is represented by basis states of remote ($r$) and near ($n$) classes and follow L\"owdin's \cite{voon2009kp, elder2011double, luttinger1955motion} procedure for treating the remote states as a perturbation to the near states. The Hamiltonian matrix can be schematically expressed as follows:
\begin{equation}
\hat{H}_{vc} = \hat H_0 + \hat H_{\{vc\} \in n} + \hat H_{\{vc\} \in r}
\end{equation}
where the last term is responsible for coupling away from $\Gamma$-point and can be considered as insignificant in direct interband absorption in the $\Gamma$ - point. The wavefunctions for the near states, on the other hand, are dictated by the point symmetry group of the crystal lattice, and usually can be expressed in terms of spherical tensors \cite{voon2009kp}. This means that one of the proper bases is the total angular momentum basis. It is important to note that the corresponding operator $\hat{H}$ should be projected onto the extended set ($| \psi_r \rangle \bigoplus |\psi_n \rangle$).

To extract the electronic near states in Total Angular Momentum (TAM) basis we propose to consider coupling to the conduction band in the standard Luttinger Hamiltonian matrix in 6-band model:

\begin{equation}
H_{\{vc\} \in n}(\pmb{k}) = 
\left(\begin{matrix}
\frac{1}{\sqrt{2}}S && \sqrt{2} R\\
\sqrt{2} Q && \sqrt{\frac{3}{2}} S\\
-\sqrt{3}{2} S^* && \sqrt{Q}\\
-\sqrt{2} R^* && \frac{1}{\sqrt{2}} S^*\\
\end{matrix}\right)\;\;\;\;
\end{equation}
where
\begin{gather*}
P = \frac{\hbar^2}{2m_0} \gamma_1 (k_x^2+k_y^2+k_z^2);\\
Q = \frac{\hbar^2}{2m_0} \gamma_2 (k_x^2+k_y^2-2k_z^2);\\
S = \frac{\hbar^2}{2m_0} 2 \sqrt{3} \gamma_3 (k_x - i k_y) k_z;\\
R = \frac{\hbar^2}{2m_0} (-\sqrt{3} \gamma_2(k_x^2 - k_y^2) + 2i \sqrt{3} \gamma_3 k_x k_y);
\end{gather*}
and $(\gamma_1, \gamma_2,\gamma_3)$ are the Luttinger parameters \cite{voon2009kp}, which are known to be $(6.85, 2.10, 2.90)$ for GaAs correspondingly. These terms are correct in proximity to $\pmb{k} = 0$, where one can work with only the near basis of the electron/hole wavefunctions. In many cases, excitonic contributions may be considerable even for vertical transitions in direct vicinity to the $\Gamma$-point . This is known not to be the case for GaAs photo-injection \cite{naka2016excitons}, so we assume no excitonic contribution in this work.

The remote contributions are accounted for in expansion coefficients $S(\pmb{k})$, which depend on the global symmetries of the crystal lattice and can be extracted from the group theory analysis. The Hamiltonian expanded in the TAM basis can be symbolically expressed as:

\begin{flalign}
p_{vc}^{\mu} =& \frac{\hbar}{m_0} \langle \alpha_c; f | \pmb{p}^{\mu} |\alpha_v; i \rangle = \frac{\hbar}{m_0} \times &&\nonumber\\ &\times \sum_{m_c, m_v} S_{c}^*(\pmb{k})S_{v}(\pmb{k}) \langle j_c m_c | \pmb{p}^{\mu} | j_v m_v \rangle
\end{flalign}
where $|\alpha_s; i \rangle$ represents the linear combination of near and remote states describing the ``exact" overall behavior of the electron anywhere on the lattice, while $| j m \rangle$ are the near $\Gamma$-point atomic-like basis states, see Ref.\cite{elder2011double} for details. This formulation allows us to extract the selection rules for the plane-wave excitation of a bulk semiconductor using the Wigner-Eckart theorem:

\onecolumngrid
\vspace{-0mm}
\begin{center}
\begin{table}[h]
  \centering
  \caption{Photoelectron polarization for small $\theta_q$ (with accuracy ($O(\theta_q^2$)) for different sets of photon quantum numbers as a function of the parameter $x=2\pi b/ \lambda$. Top two rows are for unstrained GaAs and bottom two rows are  for strained GaAs.}
  \begin{tabular}{|P{1.cm}|P{3.cm}|P{3.cm}| P{3.cm}|P{3.cm}| P{3.cm}|}
    \hline
    \diagbox{$\sigma$}{$m_{\gamma}$} & $-2$ & $-1$ & $0$ & $1$ & $2$ \\ \hline
    1 & $\frac{36-x^4}{72 + 36 x^2+ 2x^4}$ & $\frac{4-x^4}{2(4 + 8x^2 + x^4)}$ & $-\frac{x^2}{4+2x^2}$ & $-1/2$ & $-1/2$ \\ \hline
    $-1$ & 1/2 & 1/2 & $\frac{x^2}{4+2x^2}$ & $-\frac{4-x^4}{2(4 + 8x^2 + x^4)}$ & $-\frac{36-x^4}{72 + 36 x^2+ 2x^4}$ \\ \hline \hline
    1 & $\frac{36-x^4}{36+x^4}$ & $\frac{4-x^4}{4+x^4}$ & $-1$ & $-1$ & $-1$ \\ \hline
    $-1$ & 1 & 1 & 1 & $-\frac{4-x^4}{4+x^4}$ & $-\frac{36-x^4}{36+x^4}$ \\ \hline
  \end{tabular}
  \label{tab1}
\end{table}
\end{center}
\twocolumngrid

\begin{equation}
p_{vc}^{\mu} = \frac{\hbar}{m_0} \sum_{m_c, m_v} \frac{C_{j_v m_v j m}^{j_c m_c}}{\sqrt{2j_f+1}} \langle j_c || p_{cv} || j_v \rangle S_{c}^*(\pmb{k})S_{v}(\pmb{k})
\label{09/08/2018_3}
\end{equation}
This way coupling of the photon TAM state to the state of the valence electron being excited becomes explicitly controlled through the Clebsch-Gordan coefficients.

Going back to the eqn. \eqref{05/18/2018_1}, one can rewrite it as:
\begin{flalign}
M_{vc}^{\mu} =& \frac{\hbar}{m_0} \sum_{\ell} e^{i(\pmb{k}_v - \pmb{k}_c + \pmb{q}) \cdot \pmb{R}_{\ell}} \sum_{m_c=-j_c}^{j_c} \sum_{m_v = -j_v}^{j_v} \frac{C_{j_v m_v j m}^{j_c m_c}}{\sqrt{2j_f+1}} \times &&\nonumber\\ &\times  (\epsilon_{\pmb{q} \lambda})_{\mu} \langle j_c || p_{cv} || j_v \rangle S_{c}^*(\pmb{k})S_{v}(\pmb{k})
\label{05/21/2018_3}
\end{flalign}
One can see, that in this form coupling of the photon helicity to the electronic TAM is explicitly extracted via Wigner-Eckart theorem.

\section{Photo-absorption of an OAM photon state} \label{subsec.IIc}

We use Bessel modes to describe the twisted photon states \cite{Afanasev:2013kaa}. The corresponding photon vector potential can be obtained by angular spectrum decomposition into plane waves with the fixed longitudinal wave vector $\vec{q}_z$ and pitch angle $\theta_q = \arctan(|\vec{q}_{\perp}|/q_z)$:
\begin{equation}
\begin{split}
\mathcal{A}_{\kappa m_{\gamma} q_z \Lambda}^{\mu}(\vec{r}, t) = e^{-i\omega t} \;\;\;\;\;\;\;\;\;\;\;\;\;\;\;\;\;\;\;\;\;\;\;\;\;\\ \int \frac{d^2 q_{\perp}}{(2\pi)^2} a_{\kappa m_{\gamma}}(\pmb{q}_{\perp}) \epsilon_{\pmb{q} \sigma}^{\mu} e^{i\pmb{q} \cdot (\pmb{\rho}-\pmb{b})}
\label{04/25/17_1}
\end{split}
\end{equation}
where $a_{\kappa m_{\gamma}}(\pmb{q}_{\perp}) = A (-i)^{m_{\gamma}}\delta (\kappa - q_{\perp}) \exp(im_{\gamma} \phi_q)$ is the Fourier amplitude.
The global polarization basis can be expressed as
\begin{equation}
\epsilon_{\pmb{q} \sigma}^{\mu} = \sum_{m} D^{1\;*}_{\sigma m} (0, -\theta_q, -\phi_q) \eta_{m}^{\mu}
\label{07/13/17/2}
\end{equation}
in terms of local set of photon helicity states such as
\begin{equation}
\eta^{\mu}_{\pm 1} = \frac{1}{\sqrt{2}} (0, \mp 1, -i, 0);\;\;\;\;\;\eta_0^{\mu} = (0,0,0,1).
\end{equation}
where $\sigma=\pm 1$ is photon helicity and $m=\pm1,0$. For a more detailed description of the twisted photon state we refer the reader to, e.g., \cite{Afanasev:2013kaa, Scholz2014, afanasev2016high}. When substituting the plane-wave vector potential in the under-bracket of eqn. \eqref{05/18/2018_2} with eqn. \eqref{04/25/17_1} for a twisted beam, we get
\begin{flalign}
M_{vc}^{\mu} =& \int \frac{d^2 q_{\perp}}{(2\pi)} a_{\kappa m_{\gamma}}(\pmb{q}_{\perp}) e^{-i\pmb{q}_{\perp} \cdot \pmb{b}_{\ell}} \int_V u_{k_c} (\pmb{r}) \times &&\nonumber \\ &\times e^{-i \pmb{k}_c \cdot \pmb{r}} e^{i \pmb{q} \cdot \pmb{r}} (\hat{\varepsilon}_{\pmb{q} \sigma} \cdot \pmb{p}) u_{k_v} (\pmb{r}) e^{i \pmb{k}_v \cdot \pmb{r}} d^3 r
\end{flalign}
where the photon's transverse wave vector is $\kappa = \sqrt{q_x^2+q_y^2}$. Distinctively from the plane-wave case, $\pmb{q} = \pmb{q}(\phi_q, \rho, z)$ cannot be assumed to be aligned with any particular direction, but instead form a cone with a fixed opening half-angle $\theta_q$ around the beam axis (for the considered Bessel mode). The reference frame is intuitively chosen to be at the optical axis of the beam.

To evaluate photoelectron excitation rates and spin polarization, we can proceed further in the following two steps. 1. Calculate transition matrix elements at   $\Gamma$-point and limit ourselves to the case when electrons in the conduction band can be assumed to be quasi-localized at a particular ion $\pmb{b}_{\ell} \approx \pmb{R}_{\ell}$, or within an elementary cell of a crystal; 2. Consider the possibility for electrons to migrate $\pmb{b}_{\ell} \neq \pmb{R}_{\ell}$, in which case OAM will be transferred from the photon to the electron remote states, as well as ionic near states.

In what follows we will choose the first step and discuss the second one at the end. Making the same approximations as before and equating $\pmb R_{\ell}$ and $\pmb{b}_{\ell}$ we get

\begin{flalign}
M_{vc}^{\mu} =& \sum_{\ell} e^{i (\pmb{k}_v-\pmb{k}_c) \cdot \pmb{R}_{\ell}} \int \frac{d^2 q_{\perp}}{(2\pi)^2} a_{\kappa m_{\gamma}}(\pmb{q}_{\perp}) e^{i\pmb{q}_{\perp} \cdot \pmb{R}_{\ell}} \times &&\nonumber \\ &\times  \int_V u_{k_c} (\pmb{r}) e^{i \pmb{q} \cdot \pmb{r}} (\hat{\varepsilon}_{\pmb{q} \sigma} \cdot \pmb{p}) u_{k_v} (\pmb{r}) d^3 r
\end{flalign}
This result is similar to the one before eqn. (2) in \cite{quinteiro2009theory}, except we keep the $(\pmb{q} \cdot \pmb{r})$-dependent exponent inside the integral and use the expansion in vector spherical harmonics

\onecolumngrid
\begin{center}
\begin{figure}[h]
\centering
\includegraphics[scale=0.35]{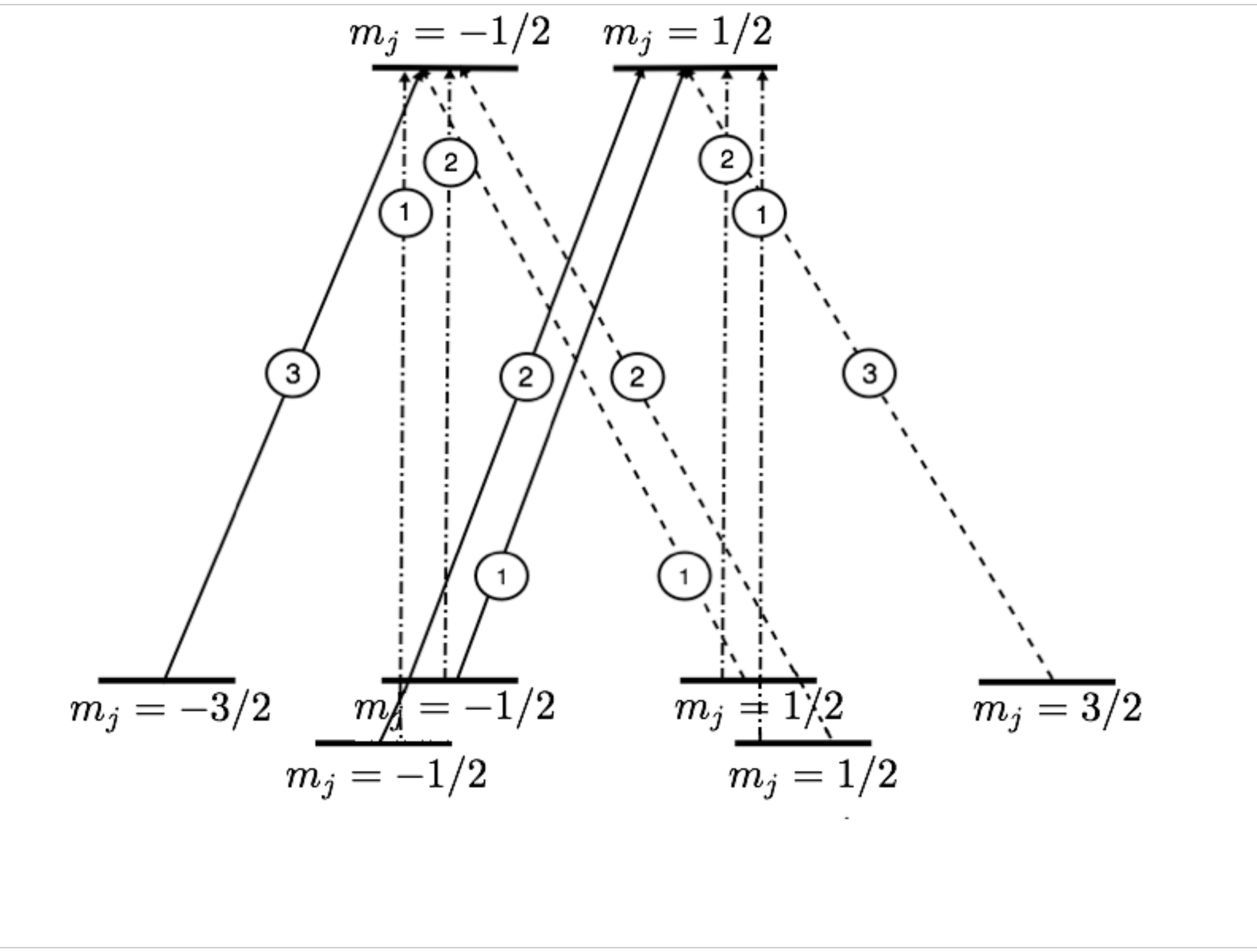}\\
\vspace{-14mm}
\caption{Energy bands for GaAs, shown along with the transitions allowed for the twisted photons with right circular polarization. Numbers in the circles indicate squares of Clebsch-Gordan coefficients. In addition to transitions with $\Delta m=1$ (solid vertical lines), $c.f.$ Ref.\cite{pierce1976photoemission}, for the twisted photons the transitions with $\Delta m=0$ (dash-dottes lines) and $\Delta m=-1$ (dashed) are also allowed. For strained GaAs, only the transitions from m=$\pm 3/2$ sublevels can be selected by properly adjusting the laser's wavelength \cite{Maruyama91}.}
\label{EnergyBands}
\end{figure}
\end{center}
\twocolumngrid

\begin{flalign}
\hat{\varepsilon}_{\pmb{q} \sigma} e^{i \pmb{q} \cdot \pmb{r}} =& -\sqrt{4\pi} \sum_{jm} \sqrt{\frac{2j+1}{2}} D_{\sigma m}^{j\;*} (0, -\phi_q, -\theta_q) \times &&\nonumber \\ &\times \sigma^{\eta+1} i^{j+\eta} A_{j\eta}^{\mu} (\pmb{q}, \pmb{r})
\end{flalign}
where $A_{j \eta}^{\mu} (\pmb{q}, \pmb{r})$ is the vector potential of multipolarity $\eta$ and order $j$. Substituting into the expression for the transition matrix element, we obtain

\begin{flalign}
M_{vc}^{\mu} =& -\sqrt{4\pi} \sum_{jm} \sqrt{2j+1}d_{\sigma m}^{j} (\theta_q) \sigma^{\eta + 1} i^{\eta+j} \times &&\nonumber \\ &\times \sum_{\ell} e^{i (\pmb{k}_v-\pmb{k}_c) \cdot \pmb{R}_{\ell}} \int \frac{d^2 q_{\perp}}{(2\pi)^2} a_{\kappa m_{\gamma}}(\pmb{q}_{\perp}) e^{i\pmb{q}_{\perp} \cdot \pmb{R}_{\ell}} e^{im\phi_q} \times &&\nonumber \\ \times & \int_V u_{k_c} (\pmb{r}) (\pmb A_{j \eta} (\pmb{q}, \pmb{r}) \cdot \pmb{p}) u_{k_v} (\pmb{r}) d^3 r
\label{05/21/2018_1}
\end{flalign}

If one would work out the dipolar approximation explicitly, for $r \sim a_0$ one will get:
\begin{equation}
A^{\mu}_{E1} \approx \frac{1}{\sqrt{6\pi}} j_0 (qr) \sum_{i} \hat{\chi}_{\mu}^i
\end{equation}
where $i$ is the summation over chiral polarization basis. After substituting it in eqn. \eqref{05/21/2018_1} one gets:
\begin{flalign}
M_{vc}^{\mu} =& -\sqrt{\frac{2}{3}} \sum_{jm} d_{\sigma m}^{j} (\theta_q) \sigma^{\eta + 1} i^{\eta+j} \sum_{\ell} e^{i (\pmb{k}_v-\pmb{k}_c) \cdot \pmb{R}_{\ell}} \times &&\nonumber \\ \times &\int \frac{d^2 q_{\perp}}{(2\pi)^2} a_{\kappa m_{\gamma}}(\pmb{q}_{\perp}) e^{i\pmb{q}_{\perp} \cdot \pmb{R}_{\ell}} j_{0}(qR_{\ell}) e^{im\phi_q} \times &&\nonumber \\ \times & \sum_{i} \chi_{\mu}^{i} \int_V u_{k_c} (\pmb{r}) p^{\mu} u_{k_v} (\pmb{r}) d^3 r
\label{05/21/2018_2}
\end{flalign}
where we are following the classical treatment for now and assuming \eqref{08/09/2018_4} as before. The last integral is the matrix element that one extracts from conventional methods of solving for the semiconductor band structure. We can now factorize and calculate the integral over the reciprocal space using the vertical transition approximation:
\begin{flalign}
\int \frac{d^2 q_{\perp}}{(2\pi)^2} & a_{\kappa m_{\gamma}}(\pmb{q}_{\perp}) e^{i\pmb{q}_{\perp} \cdot \pmb{R}_{\ell}} j_{0}(qR_{\ell}) e^{im\phi_q} = &&\nonumber \\ & (-i)^{2m_{\gamma} - m} \sqrt{\frac{\kappa}{2\pi}} j_0 (q' R_{\ell}) J_{m_{\gamma} - m}(\kappa R_{\ell}^{\perp})
\end{flalign}
where $q' = \sqrt{\kappa^2+q_z^2}$. Substituting it into the eqn. \eqref{05/21/2018_2} and using the form of the matrix element eqn. \eqref{05/21/2018_3}, we
get
\begin{flalign}
M_{vc}^{\mu} =& - \sqrt{\frac{\kappa}{3\pi}} \frac{\hbar}{m_0} \sum_{jm} (-i)^{2m_{\gamma}-m} d_{\sigma m}^{j} (\theta_q) \sigma^{\eta + 1} i^{\eta+j} \times &&\nonumber \\ &\times \sum_{\ell} e^{i (\pmb{k}_v-\pmb{k}_c) \cdot \pmb{R}_{\ell}} j_0 (q'R_{\ell}) J_{m_{\gamma}-m} (\kappa R_{\ell}^{\perp}) \times &&\nonumber \\ &\times   \sum_{m_c, m_v }\sqrt{\frac{2j+1}{2j_f+1}} C_{j_v m_v j m}^{j_c m_c} \langle j_c || \pmb{p}^{\mu} || j_v \rangle S_{c}^*(\pmb{k})S_{v}(\pmb{k})
\label{05/21/2018_5}
\end{flalign}

As one can see from this equation, c.f. Ref.\cite{afanasev2016high}, in this case the quantum selection rules coming from the symmetry of the individual ions of the lattice are not affected. The momentum transfer to the crystal lattice is described by the sum

\begin{equation}
\sum_{\ell} e^{i (\pmb{k}_v-\pmb{k}_c) \cdot \pmb{R}_{\ell}} j_0 (q'R_{\ell}) J_{m_{\gamma}-m} (\kappa R_{\ell}^{\perp})
\label{05/21/2018_4}
\end{equation}

\onecolumngrid
\begin{center}
\begin{figure}[h]
\centering
\includegraphics[scale=0.6]{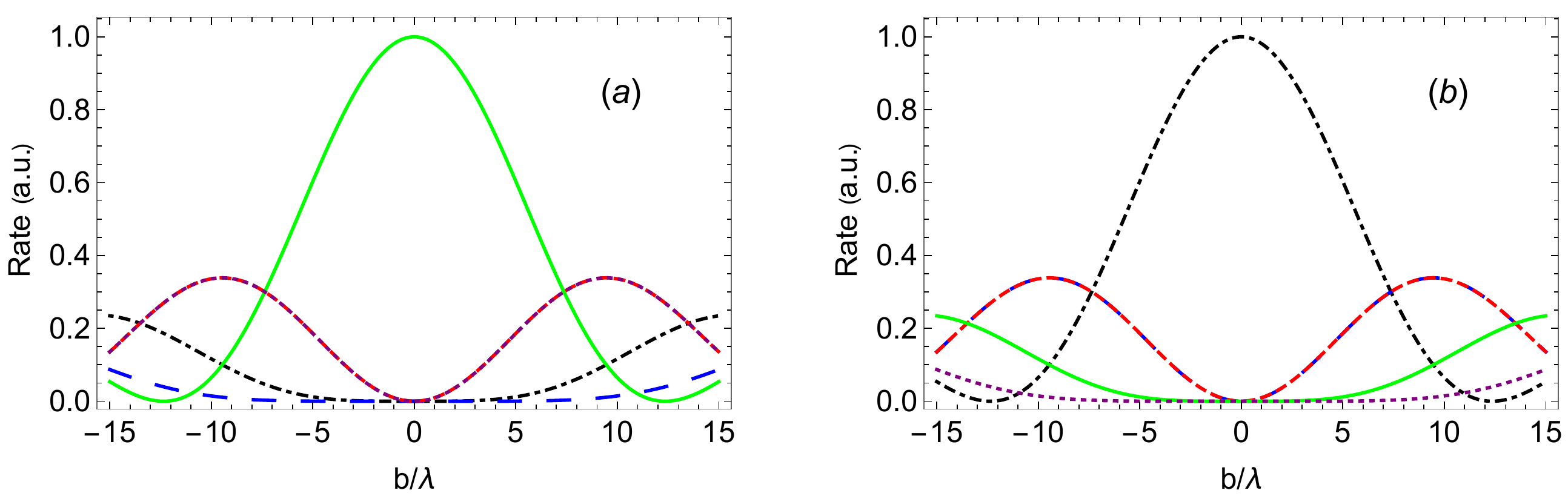}
\vspace{-1mm}
\caption{Photo-exitation rate as a function of electron's distance to the optical vortex center for the vortex pitch angle $\theta_q = 0.1$~rad; long-dashed blue is $m_{\gamma} = -2$, dot-dashed black is $m_{\gamma} = -1$, dashed red is $m_{\gamma} = 0$, solid green is $m_{\gamma} = 1$, dotted purple is $m_{\gamma} = 2$. Incoming photon polarization is (a) -- left-circular, (b) -- right-circular.}
\label{08/09/2018/1}
\end{figure}
\end{center}
\twocolumngrid

An important observation to make is that in eqn. \eqref{05/21/2018_5} we have three factorized contributions: angular, envelope and ionic. Angular contribution $d_{\sigma m}^{j} (\theta_q)$ and the envelope sum \eqref{05/21/2018_4} are the terms, analogous to the ones in eqn. (13) in \cite{afanasev2016high}, responsible for the modified selection rules. Similar to the results of Refs.\cite{Scholz2014,afanasev2016high}, Wigner rotation matrices  $d_{\sigma m}^{j} (\theta_q)$ suppress transitions with $m \neq \sigma$, while Bessel function imposes the TAM projection conservation at the optical vortex center, $m_{\gamma} = m$, where $m=m_f - m_i$ is ensured by Clebsch-Gordan coefficients.
These atom-like selection rules adequately describe photo-excitations in $\Gamma$-point in GaAs \cite{pierce1976photoemission, dunham1993investigations}, correctly predicting spin polarization of photoelectrons excited into the conduction band by Gaussian beams.

Coming back to the case when the equality $\pmb{R}_{\ell} = \pmb{b}_{\ell}$ is strongly broken, meaning that the excited electrons migrate substantially from the location of its photo-excitation within the timeframe of interest (before ejection/recombination etc.), the situation changes substantially. One should explicitly calculate the lattice response, describing coupling of the photon topological charge to the near and remote states of the electrons. In this case one is not allowed to make an approximation \eqref{08/09/2018_4}, since it assumes that the electromagnetic field varies insignificantly in the photo-excited region of the crystal, meaning that optical response of all the ions is to a large degree similar. However, this condition is violated for singular beams. This development is the subject for extensive group-theoretical analysis and computational effort. It should involve singular beam modeling and the electron dynamics simulations. Though having great potential, it is left outside of the scope of this paper.

\section{Single ion approximation}

It is known in the field \cite{dunham1993investigations} that GaAs electron wavefunctions can be modelled with p-like (valence zone) and s-like (conduction zone) single atom electron wave functions. This approximation is valid for transitions near the bandgap $\pmb q \sim 0$. According to Pierce \emph{et al.} \cite{pierce1976photoemission}, ``...when $\hbar \omega \gtrsim E_g$, only transitions between states with well-defined orbital angular momentum, characteristic of $\Gamma$, are induced..." for bulk GaAs.  These relate to the formalism, developed in Sec. \ref{subsec.IIc} by neglecting the photon TAM transfer to the ion lattice in eq. \eqref{05/21/2018_5}. For electron states in $\Gamma$ - point in GaAs ($\lambda=871$nm,$\;a_0/\lambda \ll 1$) atom-like selection rules for the inter-band electron transition is known to be a good approximation \cite{dunham1993investigations} for plane-wave photons or Gaussian beams. Here we apply the modified atom-like selection rules for the case of OAM-light photo-excitations.

The energy band structure for GaAs is shown in FIG.\ref{EnergyBands}, along with new transitions allowed only for the twisted photons. It results in the transition matrix element

\begin{flalign}
M_{vc}^{\mu} =& - \sqrt{\frac{\kappa}{3\pi}} \frac{\hbar}{m_0} \sum_{jm} (-i)^{2m_{\gamma}-m - \eta-j} \sigma^{\eta + 1} \times &&\nonumber \\ &\times \underbracket{d_{\sigma m}^{j} (\theta_q) J_{m_{\gamma}-m} (\kappa R)} \sum_{m_c, m_v }\sqrt{\frac{2j+1}{2j_f+1}} \times &&\nonumber \\ &\times C_{j_v m_v j m}^{j_c m_c} \langle j_c || \pmb{p}^{\mu} || j_v \rangle
\label{10/17/2018_1}
\end{flalign}
where $R$ identifies the position of a particular bound electron with respect to the photon beam quantization axis. This is the result identical to the one for photo-excitations of single trapped ions, developed in \cite{afanasev2018experimental,afanasev2018E2M1}. It is important to note the convenient separation of the terms responsible for OAM-transfer in the under-bracket eqn. \eqref{10/17/2018_1} from the plane-wave contribution $\langle j_c || \pmb{p}^{\mu} || j_v \rangle$.

The photo-excitation rates $\sum_{m_vm_c} |M_{vc}^{\mu}|$, obtained from \eqref{10/17/2018_1} as a function of distance from the beam center are depicted in FIG. \ref{08/09/2018/1}. Since the considered transition $P_{3/2}\to S_{1/2}$ is of electric-dipole type, the rate is proportional to the beam intensity profile \cite{afanasev2016high}.

\onecolumngrid
\begin{center}
\begin{figure}[h]
\centering
\includegraphics[scale=0.6]{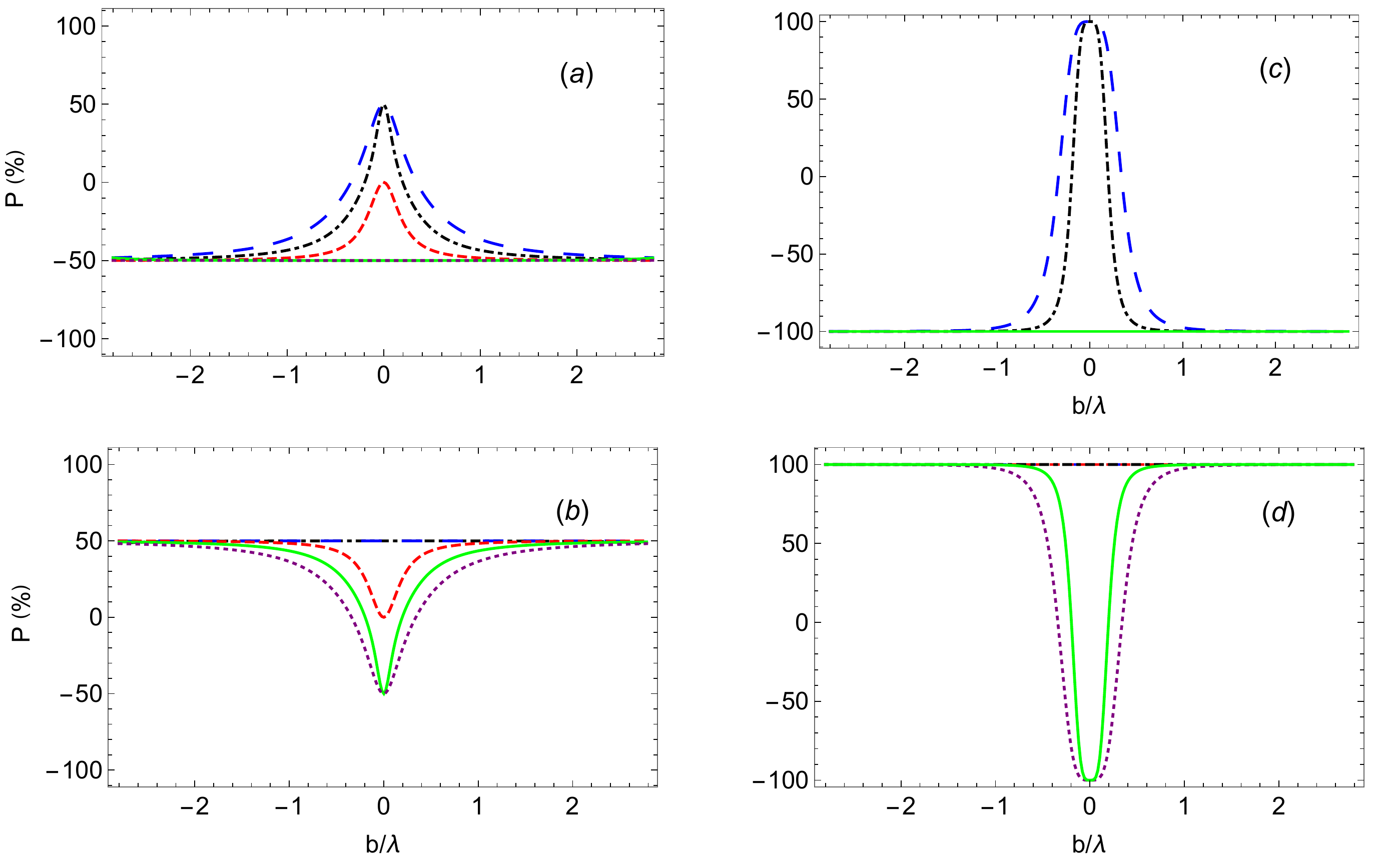}\\
\vspace{-2mm}
\caption{Photoelectron polarization as a function of the electron's distance to the optical vortex center for the vortex pitch angle $\theta_q = 0.1$~rad; long-dashed blue is $m_{\gamma} = -2$, dot-dashed black is $m_{\gamma} = -1$, dashed red is $m_{\gamma} = 0$, solid green is $m_{\gamma} = 1$, dotted purple is $m_{\gamma} = 2$. GaAs sample and incoming photon polarization, are, respectively (a) -- unstrained and left-circular, (b) -- unstrained and right-circular, (c) -- strained and left-circular polarization and (d) -- strained and right-circular.}
\label{06/26/2018/1}
\end{figure}
\end{center}
\twocolumngrid

In FIG. \ref{06/26/2018/1} we plotted the photoelectron's spin polarization in the conduction band
\begin{equation}
P = \frac{\sum_{m_v} (|M_{v;1/2}^{\mu}|^2-|M_{v;-1/2}^{\mu}|^2)}{\sum_{m_v} (|M_{v;1/2}^{\mu}|^2+|M_{v;-1/2}^{\mu}|^2)}
\end{equation}
including only transitions from $P_{3/2}$ band for unstrained GaAs (left) and only from m=$\pm 3/2$ for strained GaAs (right) ($c.f.$ Ref.\cite{pierce1976photoemission} and Ref.\cite{Maruyama91}, respectively). The polarization maxima for circularly polarized cases are narrow and localized in the low intensity regions at the center and on the periphery. One can see that the polarization remains unchanged for spin and OAM {\it aligned} with each other, while it is altered at sub-wavelength distances near the beam center for {\it anti-aligned} OAM and spin. The corresponding analytic expressions for the electron polarization for $\theta_q \rightarrow 0$ are listed in Table \ref{tab1}. It should be noted that as long as the pitch angle $\theta_q$ is small, the spatial pattern of polarization is independent of it within $O(\theta_q^2)$ accuracy. Typical twisted-photon generation approaches achieve $\theta_q\approx 0.1$ \cite{afanasev2018experimental}, making these predictions accurate at per cent level.

Crucial observation is that when averaging the electron polarization over the entire transverse beam profile, one would get $\langle P \rangle   = 0.5 \cos \theta_q$ for unstrained GaAs (and left-circularly polarized light) and twice that value for strained GaAs, which corresponds to a zero-OAM beam incident at a pitch angle $\theta_q $ to the quantization axis. The sign of position-averaged polarization is opposite for right circular polarization, and it leads to null OAM effect on photo-electron polarization from linearly-polarized light, in agreement with experiment \cite{clayburn2013search}. Similarly to the twisted-light effect on individual atoms \cite{Afanasev:2013kaa}, integration over the beam position results in the same absorption rate as for standard Gaussian beams up to an overall factor $\cos \theta_q$. Therefore, sub-wavelength position resolution is required for experimentally observing novel polarization effects due to photon's OAM.

Electron kinetics and currents generated by twisted light in conduction band were extensively studied by Quinteiro et al. \cite{quinteiro2009electric, quinteiro2009theory, quinteiro2010twisted}. Authors in \cite{cygorek2015insensitivity} further develop that approach to study spin dynamics and predict only slight differences due to the photon OAM, also in agreement with experiment \cite{clayburn2013search}. In these references, the electron response was averaged over the beam profile, see Eq.(11) of Ref.\cite{quinteiro2010twisted} and  Eq.(2) of  Ref.\cite{cygorek2015insensitivity}, that, as we show in our paper, leads to the insensitivity to optical 

\onecolumngrid
\begin{center}
\begin{figure}
\centering
\includegraphics[scale=0.6]{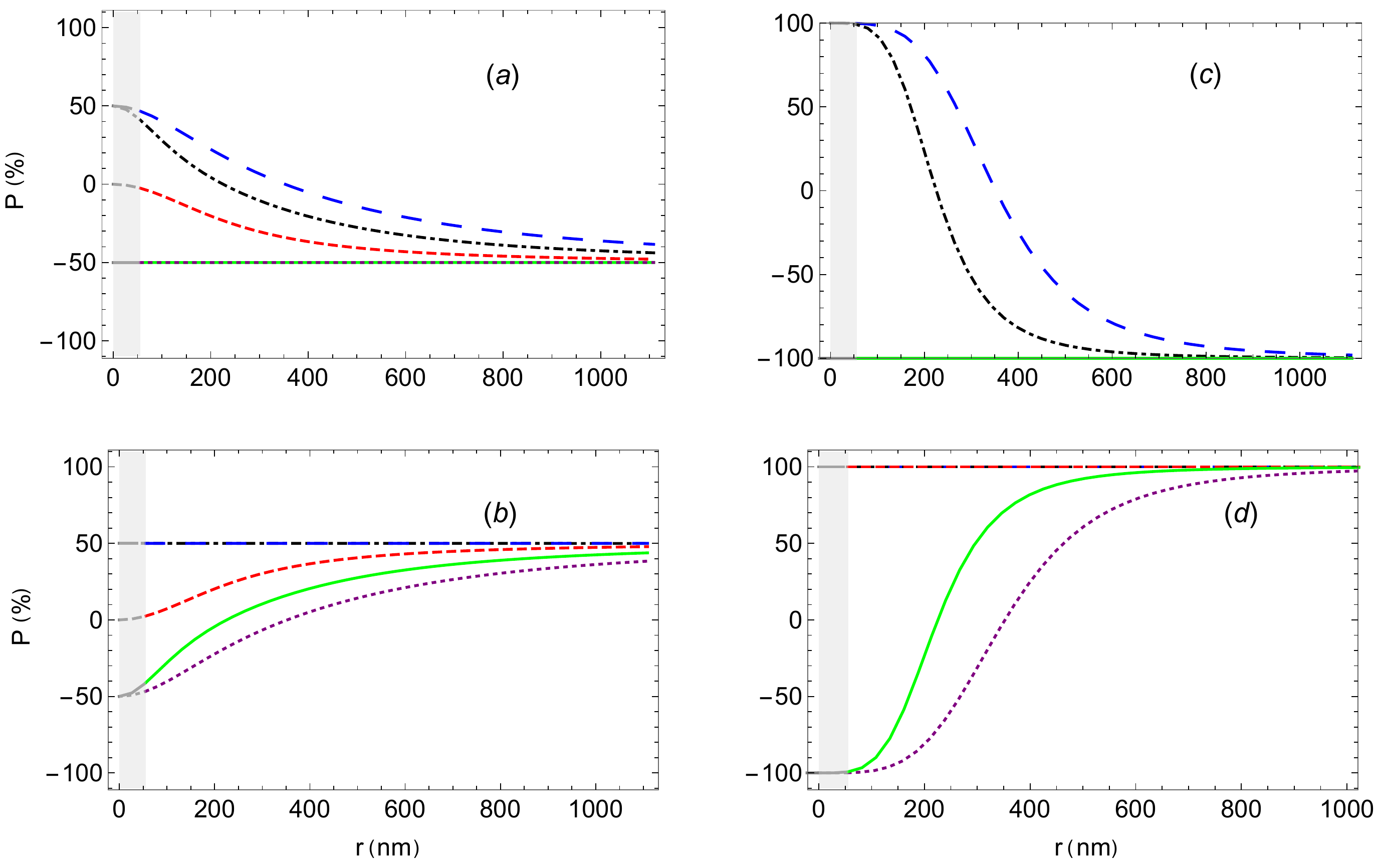}\\
\vspace{-5mm}
\caption{Photoelectron polarization averaged over the circular mesoscopic semiconducting target of radius $r$ centered at the beam axis with the pitch angle $\theta_q = 0.1$~rad; long-dashed blue is $m_{\gamma} = -2$, dot-dashed black is $m_{\gamma} = -1$, dashed red is $m_{\gamma} = 0$, solid green is $m_{\gamma} = 1$, dotted purple is $m_{\gamma} = 2$.  GaAs sample and incoming photon polarization, are, respectively (a) -- unstrained and left-circular, (b) -- unstrained and right-circular, (c) -- strained and left-circular polarization and (d) -- strained and right-circular. Shadowed vertical regions indicate 10x lattice periods for GaAs (r=5.65 nm).}
\label{08/14/2018/1}
\end{figure}
\end{center}
\twocolumngrid

\hspace{-3.5mm}OAM. The details of underlying approximations and assumptions are presented in \cite{quinteiro2009theory}. In contrast, we explore the impact parameter dependence close to the $\Gamma$-point, {\it i.e.} spacial location of the photoelectron immediately after the absorption. It is well-known that transitions in $\Gamma$-point in GaAs are well approximated by p-s atomic-like transitions. Hence, we allow OAM coupling to the electron near-states that modifies the local set of selection rules, similar to single atoms.

From the calculations and discussion above it follows that in order to observe the described polarization effects, one needs to design an experiment with sub-wavelength position resolution and a method to mitigate effects of photo-electron diffusion that could smear spatial distribution of polarization. One possibility is to use a sub-wavelength sample of GaAs. For this purpose, we calculated the average degree of photo-electron polarization in conduction band as a function of the GaAs sample size, centered at the beam's optical axis FIG. \ref{08/14/2018/1}. As one would expect, average asymmetry approaches the plane-wave (or zero-OAM) limit as the radius $r$ increases. Shaded areas correspond to the limit when dimensional confinement effects become strong, and bulk GaAs approach is no longer valid. Since confinement effects are not accounted for in this formalism, our calculations are expected to diverge from experimental results in the shaded regions. Nevertheless, one can see from FIG. \ref{08/14/2018/1} that sample sizes of $\approx$200 nm potentially allow to clearly observe the polarization effects due to OAM.

In our approach we start from L\"owdin's double-group formalism, followed by the approximate treatment GaAs electron wavefunctions as the ones of s- and p-like single-ion electron states. These allows exploiting electron OAM-conservation in the near states with subsequent description of an angular momentum coupling via the Wigner-Eckart theorem. Convenient factorization of the plane-wave contributions at the level of the matrix element has been achieved. In addition, we made specific predictions for photoelectron polarization from the twisted light that could not be found in prior literature. 

We emphasize that proposed is an atom-like model approach applicable near $\Gamma$-point, that has to be supplemented by calculations of electron spin-relaxation effects. Predicted here is an initial (in time) and localized (in position) photoelectron polarization which evolution has to be analyzed by standard methods previously developed for GaAs.

\section{Conclusions}

We have shown theoretically that twisted photon beams have the capability of transferring the angular momentum degrees of freedom to electrons in semiconductors, when exciting them in $\Gamma$-point into conduction band of semiconducting crystal lattice. The transfer signature may be observable in the form of electron polarization. It differs from the theoretical value of 50\% polarization for unstrained GaAs and 100\% for strained in beams without OAM. Spatial distribution of the photon absorption rates are predicted to be proportional to the beam intensity profile in the linear regime. As a result, the absorption rates follow doughnut-like behavior characteristic of Bessel or Laguerre-Gaussian beams. On the other hand, the photoelectron spatial polarization pattern is predicted to exhibit concentric rings of differing polarization states from those expected from beams with zero OAM. In the case with circularly polarized photons, one may expect to observe  polarization singularities near the optical vortex center for the case when photon OAM and spin are anti-aligned, as well as the rings, mentioned earlier. Spatial electron confinement in arrays of quantum dots can produce similar and even enhanced (due to higher position resolution) polarization patterns. We found that photo-electron polarization averaged over the beam position is zero for linearly polarized OAM light, in full consistency with experiment \cite{clayburn2013search}.

We demonstrated that for electron excitations in $\Gamma$ - point in GaAs, the photon beams with OAM should be able to transfer TAM directly to electron degrees of freedom. It has been argued that the traditional approach of position-averaging the optical response over the entire semiconductor will not be suitable for the more general description of photo-excitations in a bulk semiconductor by singular beams. Possible ways to tackle this problem have been discussed. We showed particular cases when OAM and spin AM completely cancel each other in their effect on electron's magnetic quantum number, leaving only electron transitions with $\Delta m=0$; or OAM completely dominating photon's spin effect, such that $\Delta m$ has the opposite sign compared to the transitions caused by plane-wave photons. 

We show that the geometric dimensions of the areas of the abnormal photoelectron polarization in vortex center is only defined by the beam's topological charge and photons wavelength in paraxial approximation. It is shown to be independent of the vortex pitch angle $\theta_q$ or the ``doughnut" radius in the Bessel-beam intensity profile, while the photoexcitation rate remains strongly dependent on  $\theta_q$ . The electron's polarization significantly changes over the sub-wavelength distance ($\approx\lambda/3$) away from the vortex center. Therefore the mentioned independence of polarization profile on the ``doughnut" size can be instrumental for polarization-enhanced sub-wavelength resolution optical microscopy. The same polarization feature can possibly be used for alignment of laser beams in space with sub-wavelength accuracy.

To verify the above predictions experimentally, it is important to localize the photoexcited electron within photon's sub-wavelength distance around the optical vortex center. Otherwise, the electron's diffusion would smear the polarization pattern. Such effects are likely to be seen in sub-micron-sized samples of GaAs, or quantum dots, with non-overlapping (isolated) atomic-like electron states. 

If verified experimentally, the predicted polarization behavior could contribute to the study of semiconductor electron devices, polarized electron sources, and photovoltaic cells. It will provide new insight into spatial-temporal dynamics of photoelectrons at the scales of a fraction of photon's wavelength. This formalism can be directly applied to the development of fluorescence theory in GaAs, which potentially could help us learn more about its bulk properties and band structure. Using the formulation of the photo-absorption matrix element, from section \ref{subsec.IIc}, one could also proceed with the development of the solid state theory of dark excitons \cite{syouji2017creation} in semiconductors.

\section*{Acknowledgements}
AA and MS would like to acknowledge support of Gus Weiss endowment at The George Washington University. We appreciate useful discussions with members of Jefferson Lab's Photoinjector group, and with Prof. Robert Alfano. MS especially thanks Nicholas Gorgone for his help in working on this manuscript.

\bibliography{Master_b} 

\begin{thebibliography}{37}%
\makeatletter
\providecommand \@ifxundefined [1]{%
 \@ifx{#1\undefined}
}%
\providecommand \@ifnum [1]{%
 \ifnum #1\expandafter \@firstoftwo
 \else \expandafter \@secondoftwo
 \fi
}%
\providecommand \@ifx [1]{%
 \ifx #1\expandafter \@firstoftwo
 \else \expandafter \@secondoftwo
 \fi
}%
\providecommand \natexlab [1]{#1}%
\providecommand \enquote  [1]{``#1''}%
\providecommand \bibnamefont  [1]{#1}%
\providecommand \bibfnamefont [1]{#1}%
\providecommand \citenamefont [1]{#1}%
\providecommand \href@noop [0]{\@secondoftwo}%
\providecommand \href [0]{\begingroup \@sanitize@url \@href}%
\providecommand \@href[1]{\@@startlink{#1}\@@href}%
\providecommand \@@href[1]{\endgroup#1\@@endlink}%
\providecommand \@sanitize@url [0]{\catcode `\\12\catcode `\$12\catcode
  `\&12\catcode `\#12\catcode `\^12\catcode `\_12\catcode `\%12\relax}%
\providecommand \@@startlink[1]{}%
\providecommand \@@endlink[0]{}%
\providecommand \url  [0]{\begingroup\@sanitize@url \@url }%
\providecommand \@url [1]{\endgroup\@href {#1}{\urlprefix }}%
\providecommand \urlprefix  [0]{URL }%
\providecommand \Eprint [0]{\href }%
\providecommand \doibase [0]{http://dx.doi.org/}%
\providecommand \selectlanguage [0]{\@gobble}%
\providecommand \bibinfo  [0]{\@secondoftwo}%
\providecommand \bibfield  [0]{\@secondoftwo}%
\providecommand \translation [1]{[#1]}%
\providecommand \BibitemOpen [0]{}%
\providecommand \bibitemStop [0]{}%
\providecommand \bibitemNoStop [0]{.\EOS\space}%
\providecommand \EOS [0]{\spacefactor3000\relax}%
\providecommand \BibitemShut  [1]{\csname bibitem#1\endcsname}%
\let\auto@bib@innerbib\@empty
\bibitem [{\citenamefont {Allen}\ \emph {et~al.}(1992)\citenamefont {Allen},
  \citenamefont {Beijersbergen}, \citenamefont {Spreeuw},\ and\ \citenamefont
  {Woerdman}}]{Allen:1992zz}%
  \BibitemOpen
  \bibfield  {author} {\bibinfo {author} {\bibfnamefont {L.}~\bibnamefont
  {Allen}}, \bibinfo {author} {\bibfnamefont {M.}~\bibnamefont
  {Beijersbergen}}, \bibinfo {author} {\bibfnamefont {R.}~\bibnamefont
  {Spreeuw}}, \ and\ \bibinfo {author} {\bibfnamefont {J.}~\bibnamefont
  {Woerdman}},\ }\bibfield  {title} {\bibinfo {title} {Orbital angular momentum
  of light and the transformation of laguerre-gaussian laser modes},\ }\href
  {\doibase 10.1103/PhysRevA.45.8185} {\bibfield  {journal} {\bibinfo
  {journal} {Phys. Rev. A}\ }\textbf {\bibinfo {volume} {45}},\ \bibinfo
  {pages} {8185} (\bibinfo {year} {1992})}\BibitemShut {NoStop}%
\bibitem [{\citenamefont {Jantzi}\ \emph {et~al.}(2018)\citenamefont {Jantzi},
  \citenamefont {Jemison}, \citenamefont {Laux}, \citenamefont {Mullen},\ and\
  \citenamefont {Cochenour}}]{jantzi2018enhanced}%
  \BibitemOpen
  \bibfield  {author} {\bibinfo {author} {\bibfnamefont {A.}~\bibnamefont
  {Jantzi}}, \bibinfo {author} {\bibfnamefont {W.}~\bibnamefont {Jemison}},
  \bibinfo {author} {\bibfnamefont {A.}~\bibnamefont {Laux}}, \bibinfo {author}
  {\bibfnamefont {L.}~\bibnamefont {Mullen}}, \ and\ \bibinfo {author}
  {\bibfnamefont {B.}~\bibnamefont {Cochenour}},\ }\bibfield  {title} {\bibinfo
  {title} {Enhanced underwater ranging using an optical vortex},\ }\href@noop
  {} {\bibfield  {journal} {\bibinfo  {journal} {Opt. Express}\ }\textbf
  {\bibinfo {volume} {26}},\ \bibinfo {pages} {2668} (\bibinfo {year}
  {2018})}\BibitemShut {NoStop}%
\bibitem [{\citenamefont {Mamani}\ \emph {et~al.}(2018)\citenamefont {Mamani},
  \citenamefont {Shi}, \citenamefont {Ahmed}, \citenamefont {Karnik},
  \citenamefont {Rodr{\'\i}guez-Contreras}, \citenamefont {Nolan},\ and\
  \citenamefont {Alfano}}]{mamani2018transmission}%
  \BibitemOpen
  \bibfield  {author} {\bibinfo {author} {\bibfnamefont {S.}~\bibnamefont
  {Mamani}}, \bibinfo {author} {\bibfnamefont {L.}~\bibnamefont {Shi}},
  \bibinfo {author} {\bibfnamefont {T.}~\bibnamefont {Ahmed}}, \bibinfo
  {author} {\bibfnamefont {R.}~\bibnamefont {Karnik}}, \bibinfo {author}
  {\bibfnamefont {A.}~\bibnamefont {Rodr{\'\i}guez-Contreras}}, \bibinfo
  {author} {\bibfnamefont {D.}~\bibnamefont {Nolan}}, \ and\ \bibinfo {author}
  {\bibfnamefont {R.}~\bibnamefont {Alfano}},\ }\bibfield  {title} {\bibinfo
  {title} {Transmission of classically entangled beams through mouse brain
  tissue},\ }\href@noop {} {\bibfield  {journal} {\bibinfo  {journal} {J
  Biophotonics}\ ,\ \bibinfo {pages} {e201800096}} (\bibinfo {year}
  {2018})}\BibitemShut {NoStop}%
\bibitem [{\citenamefont {Clayburn}\ \emph {et~al.}(2013)\citenamefont
  {Clayburn}, \citenamefont {McCarter}, \citenamefont {Dreiling}, \citenamefont
  {Poelker}, \citenamefont {Ryan},\ and\ \citenamefont
  {Gay}}]{clayburn2013search}%
  \BibitemOpen
  \bibfield  {author} {\bibinfo {author} {\bibfnamefont {N.}~\bibnamefont
  {Clayburn}}, \bibinfo {author} {\bibfnamefont {J.}~\bibnamefont {McCarter}},
  \bibinfo {author} {\bibfnamefont {J.}~\bibnamefont {Dreiling}}, \bibinfo
  {author} {\bibfnamefont {M.}~\bibnamefont {Poelker}}, \bibinfo {author}
  {\bibfnamefont {D.}~\bibnamefont {Ryan}}, \ and\ \bibinfo {author}
  {\bibfnamefont {T.}~\bibnamefont {Gay}},\ }\bibfield  {title} {\bibinfo
  {title} {Search for spin-polarized photoemission from gaas using light with
  orbital angular momentum},\ }\href@noop {} {\bibfield  {journal} {\bibinfo
  {journal} {Phys. Rev. B}\ }\textbf {\bibinfo {volume} {87}},\ \bibinfo
  {pages} {035204} (\bibinfo {year} {2013})}\BibitemShut {NoStop}%
\bibitem [{\citenamefont {Miao}\ \emph {et~al.}(2016)\citenamefont {Miao},
  \citenamefont {Zhang}, \citenamefont {Sun}, \citenamefont {Walasik},
  \citenamefont {Longhi}, \citenamefont {Litchinitser},\ and\ \citenamefont
  {Feng}}]{miao2016orbital}%
  \BibitemOpen
  \bibfield  {author} {\bibinfo {author} {\bibfnamefont {P.}~\bibnamefont
  {Miao}}, \bibinfo {author} {\bibfnamefont {Z.}~\bibnamefont {Zhang}},
  \bibinfo {author} {\bibfnamefont {J.}~\bibnamefont {Sun}}, \bibinfo {author}
  {\bibfnamefont {W.}~\bibnamefont {Walasik}}, \bibinfo {author} {\bibfnamefont
  {S.}~\bibnamefont {Longhi}}, \bibinfo {author} {\bibfnamefont {N.~M.}\
  \bibnamefont {Litchinitser}}, \ and\ \bibinfo {author} {\bibfnamefont
  {L.}~\bibnamefont {Feng}},\ }\bibfield  {title} {\bibinfo {title} {Orbital
  angular momentum microlaser},\ }\href@noop {} {\bibfield  {journal} {\bibinfo
   {journal} {Science}\ }\textbf {\bibinfo {volume} {353}},\ \bibinfo {pages}
  {464} (\bibinfo {year} {2016})}\BibitemShut {NoStop}%
\bibitem [{\citenamefont {Yao}\ and\ \citenamefont {Padgett}(2011)}]{Yao11}%
  \BibitemOpen
  \bibfield  {author} {\bibinfo {author} {\bibfnamefont {A.}~\bibnamefont
  {Yao}}\ and\ \bibinfo {author} {\bibfnamefont {M.}~\bibnamefont {Padgett}},\
  }\bibfield  {title} {\bibinfo {title} {Orbital angular momentum: Origins,
  behavior and applications},\ }\href@noop {} {\bibfield  {journal} {\bibinfo
  {journal} {Adv. Opt. Photon.}\ }\textbf {\bibinfo {volume} {3}},\ \bibinfo
  {pages} {161} (\bibinfo {year} {2011})}\BibitemShut {NoStop}%
\bibitem [{\citenamefont {Torres}\ and\ \citenamefont
  {Torner}(2011)}]{torres2011twisted}%
  \BibitemOpen
  \bibfield  {author} {\bibinfo {author} {\bibfnamefont {J.~P.}\ \bibnamefont
  {Torres}}\ and\ \bibinfo {author} {\bibfnamefont {L.}~\bibnamefont
  {Torner}},\ }\href@noop {} {\bibinfo {title} {Twisted photons: applications
  of light with orbital angular momentum}}\ (\bibinfo  {publisher} {John Wiley
  \& Sons},\ \bibinfo {year} {2011})\BibitemShut {NoStop}%
\bibitem [{\citenamefont {Andrews}(2011)}]{andrews2011structured}%
  \BibitemOpen
  \bibfield  {author} {\bibinfo {author} {\bibfnamefont {D.~L.}\ \bibnamefont
  {Andrews}},\ }\href@noop {} {\bibinfo {title} {Structured light and its
  applications: An introduction to phase-structured beams and nanoscale optical
  forces}}\ (\bibinfo  {publisher} {Academic Press},\ \bibinfo {year}
  {2011})\BibitemShut {NoStop}%
\bibitem [{\citenamefont {Wisniewski-Barker}\ and\ \citenamefont
  {Padgett}(2015)}]{Padgett2015}%
  \BibitemOpen
  \bibfield  {author} {\bibinfo {author} {\bibfnamefont {E.}~\bibnamefont
  {Wisniewski-Barker}}\ and\ \bibinfo {author} {\bibfnamefont {M.}~\bibnamefont
  {Padgett}},\ }\bibfield  {title} {\bibinfo {title} {Orbital angular
  momentum},\ }\href@noop {} {\bibfield  {journal} {\bibinfo  {journal}
  {Photonics: Scientific Foundations, Technology and Applications}\ }\textbf
  {\bibinfo {volume} {1}},\ \bibinfo {pages} {321} (\bibinfo {year}
  {2015})}\BibitemShut {NoStop}%
\bibitem [{\citenamefont {Franke-Arnold}(2017)}]{Franke-Arnold2017}%
  \BibitemOpen
  \bibfield  {author} {\bibinfo {author} {\bibfnamefont {S.}~\bibnamefont
  {Franke-Arnold}},\ }\bibfield  {title} {\bibinfo {title} {Optical angular
  momentum and atoms},\ }\href
  {http://rsta.royalsocietypublishing.org/content/375/2087/20150435} {\bibfield
   {journal} {\bibinfo  {journal} {Phil. Trans. R. Soc. A}\ }\textbf {\bibinfo
  {volume} {375}} (\bibinfo {year} {2017})}\BibitemShut {NoStop}%
\bibitem [{\citenamefont {Padgett}(2017)}]{padgett2017orbital}%
  \BibitemOpen
  \bibfield  {author} {\bibinfo {author} {\bibfnamefont {M.~J.}\ \bibnamefont
  {Padgett}},\ }\bibfield  {title} {\bibinfo {title} {Orbital angular momentum
  25 years on},\ }\href@noop {} {\bibfield  {journal} {\bibinfo  {journal}
  {Opt. Express}\ }\textbf {\bibinfo {volume} {25}},\ \bibinfo {pages} {11265}
  (\bibinfo {year} {2017})}\BibitemShut {NoStop}%
\bibitem [{\citenamefont {Hernandez-Garcia}\ \emph {et~al.}(2008)\citenamefont
  {Hernandez-Garcia}, \citenamefont {Stutzman},\ and\ \citenamefont
  {O'Shea}}]{HernandezGarcia:2008zz}%
  \BibitemOpen
  \bibfield  {author} {\bibinfo {author} {\bibfnamefont {C.}~\bibnamefont
  {Hernandez-Garcia}}, \bibinfo {author} {\bibfnamefont {M.~L.}\ \bibnamefont
  {Stutzman}}, \ and\ \bibinfo {author} {\bibfnamefont {P.~G.}\ \bibnamefont
  {O'Shea}},\ }\bibfield  {title} {\bibinfo {title} {{Electron sources for
  accelerators}},\ }\href {\doibase 10.1063/1.2883909} {\bibfield  {journal}
  {\bibinfo  {journal} {Phys. Today}\ }\textbf {\bibinfo {volume} {61N2}},\
  \bibinfo {pages} {44} (\bibinfo {year} {2008})}\BibitemShut {NoStop}%
\bibitem [{\citenamefont {Pierce}\ and\ \citenamefont
  {Meier}(1976)}]{pierce1976photoemission}%
  \BibitemOpen
  \bibfield  {author} {\bibinfo {author} {\bibfnamefont {D.~T.}\ \bibnamefont
  {Pierce}}\ and\ \bibinfo {author} {\bibfnamefont {F.}~\bibnamefont {Meier}},\
  }\bibfield  {title} {\bibinfo {title} {Photoemission of spin-polarized
  electrons from {GaAs}},\ }\href@noop {} {\bibfield  {journal} {\bibinfo
  {journal} {Phys. Rev. B}\ }\textbf {\bibinfo {volume} {13}},\ \bibinfo
  {pages} {5484} (\bibinfo {year} {1976})}\BibitemShut {NoStop}%
\bibitem [{\citenamefont {Maruyama}\ \emph {et~al.}(1991)\citenamefont
  {Maruyama}, \citenamefont {Garwin}, \citenamefont {Prepost}, \citenamefont
  {Zapalac}, \citenamefont {Smith},\ and\ \citenamefont {Walker}}]{Maruyama91}%
  \BibitemOpen
  \bibfield  {author} {\bibinfo {author} {\bibfnamefont {T.}~\bibnamefont
  {Maruyama}}, \bibinfo {author} {\bibfnamefont {E.~L.}\ \bibnamefont
  {Garwin}}, \bibinfo {author} {\bibfnamefont {R.}~\bibnamefont {Prepost}},
  \bibinfo {author} {\bibfnamefont {G.~H.}\ \bibnamefont {Zapalac}}, \bibinfo
  {author} {\bibfnamefont {J.~S.}\ \bibnamefont {Smith}}, \ and\ \bibinfo
  {author} {\bibfnamefont {J.~D.}\ \bibnamefont {Walker}},\ }\bibfield  {title}
  {\bibinfo {title} {Observation of strain-enhanced electron-spin polarization
  in photoemission from {InGaAs}},\ }\href {\doibase
  10.1103/PhysRevLett.66.2376} {\bibfield  {journal} {\bibinfo  {journal}
  {Phys. Rev. Lett.}\ }\textbf {\bibinfo {volume} {66}},\ \bibinfo {pages}
  {2376} (\bibinfo {year} {1991})}\BibitemShut {NoStop}%
\bibitem [{\citenamefont {Quinteiro}\ and\ \citenamefont
  {Berakdar}(2009)}]{quinteiro2009electric}%
  \BibitemOpen
  \bibfield  {author} {\bibinfo {author} {\bibfnamefont {G.}~\bibnamefont
  {Quinteiro}}\ and\ \bibinfo {author} {\bibfnamefont {J.}~\bibnamefont
  {Berakdar}},\ }\bibfield  {title} {\bibinfo {title} {Electric currents
  induced by twisted light in quantum rings},\ }\href@noop {} {\bibfield
  {journal} {\bibinfo  {journal} {Opt. Express}\ }\textbf {\bibinfo {volume}
  {17}},\ \bibinfo {pages} {20465} (\bibinfo {year} {2009})}\BibitemShut
  {NoStop}%
\bibitem [{\citenamefont {Quinteiro}\ and\ \citenamefont
  {Tamborenea}(2009)}]{quinteiro2009theory}%
  \BibitemOpen
  \bibfield  {author} {\bibinfo {author} {\bibfnamefont {G.}~\bibnamefont
  {Quinteiro}}\ and\ \bibinfo {author} {\bibfnamefont {P.}~\bibnamefont
  {Tamborenea}},\ }\bibfield  {title} {\bibinfo {title} {Theory of the optical
  absorption of light carrying orbital angular momentum by semiconductors},\
  }\href@noop {} {\bibfield  {journal} {\bibinfo  {journal} {EPL (Europhysics
  Letters)}\ }\textbf {\bibinfo {volume} {85}},\ \bibinfo {pages} {47001}
  (\bibinfo {year} {2009})}\BibitemShut {NoStop}%
\bibitem [{\citenamefont {Quinteiro}\ and\ \citenamefont
  {Tamborenea}(2010)}]{quinteiro2010twisted}%
  \BibitemOpen
  \bibfield  {author} {\bibinfo {author} {\bibfnamefont {G.}~\bibnamefont
  {Quinteiro}}\ and\ \bibinfo {author} {\bibfnamefont {P.}~\bibnamefont
  {Tamborenea}},\ }\bibfield  {title} {\bibinfo {title} {Twisted-light-induced
  optical transitions in semiconductors: Free-carrier quantum kinetics},\
  }\href@noop {} {\bibfield  {journal} {\bibinfo  {journal} {Phys. Rev. B}\
  }\textbf {\bibinfo {volume} {82}},\ \bibinfo {pages} {125207} (\bibinfo
  {year} {2010})}\BibitemShut {NoStop}%
\bibitem [{\citenamefont {Cygorek}\ \emph {et~al.}(2015)\citenamefont
  {Cygorek}, \citenamefont {Tamborenea},\ and\ \citenamefont
  {Axt}}]{cygorek2015insensitivity}%
  \BibitemOpen
  \bibfield  {author} {\bibinfo {author} {\bibfnamefont {M.}~\bibnamefont
  {Cygorek}}, \bibinfo {author} {\bibfnamefont {P.~I.}\ \bibnamefont
  {Tamborenea}}, \ and\ \bibinfo {author} {\bibfnamefont {V.}~\bibnamefont
  {Axt}},\ }\bibfield  {title} {\bibinfo {title} {Insensitivity of spin
  dynamics to the orbital angular momentum transferred from twisted light to
  extended semiconductors},\ }\href@noop {} {\bibfield  {journal} {\bibinfo
  {journal} {Phys. Rev. B}\ }\textbf {\bibinfo {volume} {92}},\ \bibinfo
  {pages} {115301} (\bibinfo {year} {2015})}\BibitemShut {NoStop}%
\bibitem [{\citenamefont {Voon}\ and\ \citenamefont
  {Willatzen}(2009)}]{voon2009kp}%
  \BibitemOpen
  \bibfield  {author} {\bibinfo {author} {\bibfnamefont {L.~C. L.~Y.}\
  \bibnamefont {Voon}}\ and\ \bibinfo {author} {\bibfnamefont {M.}~\bibnamefont
  {Willatzen}},\ }\href@noop {} {\bibinfo {title} {The kp method: electronic
  properties of semiconductors}}\ (\bibinfo  {publisher} {Springer Science \&
  Business Media},\ \bibinfo {year} {2009})\BibitemShut {NoStop}%
\bibitem [{\citenamefont {Elder}\ \emph {et~al.}(2011)\citenamefont {Elder},
  \citenamefont {Ward}, \citenamefont {Zhang} \emph
  {et~al.}}]{elder2011double}%
  \BibitemOpen
  \bibfield  {author} {\bibinfo {author} {\bibfnamefont {W.~J.}\ \bibnamefont
  {Elder}}, \bibinfo {author} {\bibfnamefont {R.~M.}\ \bibnamefont {Ward}},
  \bibinfo {author} {\bibfnamefont {J.}~\bibnamefont {Zhang}},  \emph
  {et~al.},\ }\bibfield  {title} {\bibinfo {title} {Double-group formulation of
  k{\textperiodcentered} p theory for cubic crystals},\ }\href@noop {}
  {\bibfield  {journal} {\bibinfo  {journal} {Phys. Rev. B}\ }\textbf {\bibinfo
  {volume} {83}},\ \bibinfo {pages} {165210} (\bibinfo {year}
  {2011})}\BibitemShut {NoStop}%
\bibitem [{\citenamefont {Luttinger}\ and\ \citenamefont
  {Kohn}(1955)}]{luttinger1955motion}%
  \BibitemOpen
  \bibfield  {author} {\bibinfo {author} {\bibfnamefont {J.~M.}\ \bibnamefont
  {Luttinger}}\ and\ \bibinfo {author} {\bibfnamefont {W.}~\bibnamefont
  {Kohn}},\ }\bibfield  {title} {\bibinfo {title} {Motion of electrons and
  holes in perturbed periodic fields},\ }\href@noop {} {\bibfield  {journal}
  {\bibinfo  {journal} {Phys. Rev.}\ }\textbf {\bibinfo {volume} {97}},\
  \bibinfo {pages} {869} (\bibinfo {year} {1955})}\BibitemShut {NoStop}%
\bibitem [{\citenamefont {Afanasev}\ \emph
  {et~al.}(2013{\natexlab{a}})\citenamefont {Afanasev}, \citenamefont
  {Carlson},\ and\ \citenamefont {Mukherjee}}]{Afanasev2013kaa}%
  \BibitemOpen
  \bibfield  {author} {\bibinfo {author} {\bibfnamefont {A.}~\bibnamefont
  {Afanasev}}, \bibinfo {author} {\bibfnamefont {C.~E.}\ \bibnamefont
  {Carlson}}, \ and\ \bibinfo {author} {\bibfnamefont {A.}~\bibnamefont
  {Mukherjee}},\ }\bibfield  {title} {\bibinfo {title} {Off-axis excitation of
  hydrogenlike atoms by twisted photons},\ }\href {\doibase
  10.1103/PhysRevA.88.033841} {\bibfield  {journal} {\bibinfo  {journal} {Phys.
  Rev. A}\ }\textbf {\bibinfo {volume} {88}},\ \bibinfo {pages} {033841}
  (\bibinfo {year} {2013}{\natexlab{a}})}\BibitemShut {NoStop}%
\bibitem [{\citenamefont {Rodrigues}\ \emph {et~al.}(2016)\citenamefont
  {Rodrigues}, \citenamefont {Marcassa},\ and\ \citenamefont
  {Mendon{\c{c}}a}}]{rodrigues2016excitation}%
  \BibitemOpen
  \bibfield  {author} {\bibinfo {author} {\bibfnamefont {J.}~\bibnamefont
  {Rodrigues}}, \bibinfo {author} {\bibfnamefont {L.}~\bibnamefont {Marcassa}},
  \ and\ \bibinfo {author} {\bibfnamefont {J.}~\bibnamefont {Mendon{\c{c}}a}},\
  }\bibfield  {title} {\bibinfo {title} {Excitation of high orbital angular
  momentum rydberg states with laguerre--gauss beams},\ }\href@noop {}
  {\bibfield  {journal} {\bibinfo  {journal} {Journal of Physics B: Atomic,
  Molecular and Optical Physics}\ }\textbf {\bibinfo {volume} {49}},\ \bibinfo
  {pages} {074007} (\bibinfo {year} {2016})}\BibitemShut {NoStop}%
\bibitem [{\citenamefont {Peshkov}\ \emph {et~al.}(2017)\citenamefont
  {Peshkov}, \citenamefont {Seipt}, \citenamefont {Surzhykov},\ and\
  \citenamefont {Fritzsche}}]{peshkov2017photoexcitation}%
  \BibitemOpen
  \bibfield  {author} {\bibinfo {author} {\bibfnamefont {A.}~\bibnamefont
  {Peshkov}}, \bibinfo {author} {\bibfnamefont {D.}~\bibnamefont {Seipt}},
  \bibinfo {author} {\bibfnamefont {A.}~\bibnamefont {Surzhykov}}, \ and\
  \bibinfo {author} {\bibfnamefont {S.}~\bibnamefont {Fritzsche}},\ }\bibfield
  {title} {\bibinfo {title} {Photoexcitation of atoms by laguerre-gaussian
  beams},\ }\href@noop {} {\bibfield  {journal} {\bibinfo  {journal} {Phys.
  Rev. A}\ }\textbf {\bibinfo {volume} {96}},\ \bibinfo {pages} {023407}
  (\bibinfo {year} {2017})}\BibitemShut {NoStop}%
\bibitem [{\citenamefont {Afanasev}\ \emph
  {et~al.}(2018{\natexlab{a}})\citenamefont {Afanasev}, \citenamefont
  {Carlson}, \citenamefont {Schmiegelow}, \citenamefont {Schulz}, \citenamefont
  {Schmidt-Kaler},\ and\ \citenamefont {Solyanik}}]{afanasev2018experimental}%
  \BibitemOpen
  \bibfield  {author} {\bibinfo {author} {\bibfnamefont {A.}~\bibnamefont
  {Afanasev}}, \bibinfo {author} {\bibfnamefont {C.~E.}\ \bibnamefont
  {Carlson}}, \bibinfo {author} {\bibfnamefont {C.}~\bibnamefont
  {Schmiegelow}}, \bibinfo {author} {\bibfnamefont {J.}~\bibnamefont {Schulz}},
  \bibinfo {author} {\bibfnamefont {F.}~\bibnamefont {Schmidt-Kaler}}, \ and\
  \bibinfo {author} {\bibfnamefont {M.}~\bibnamefont {Solyanik}},\ }\bibfield
  {title} {\bibinfo {title} {Experimental verification of position-dependent
  angular-momentum selection rules for absorption of twisted light by a bound
  electron},\ }\href@noop {} {\bibfield  {journal} {\bibinfo  {journal} {New J.
  Phys.}\ }\textbf {\bibinfo {volume} {20}},\ \bibinfo {pages} {023032}
  (\bibinfo {year} {2018}{\natexlab{a}})}\BibitemShut {NoStop}%
\bibitem [{\citenamefont {Schmiegelow}\ and\ \citenamefont
  {Schmidt-Kaler}(2012)}]{schmiegelow2012light}%
  \BibitemOpen
  \bibfield  {author} {\bibinfo {author} {\bibfnamefont {C.~T.}\ \bibnamefont
  {Schmiegelow}}\ and\ \bibinfo {author} {\bibfnamefont {F.}~\bibnamefont
  {Schmidt-Kaler}},\ }\bibfield  {title} {\bibinfo {title} {Light with orbital
  angular momentum interacting with trapped ions},\ }\href@noop {} {\bibfield
  {journal} {\bibinfo  {journal} {Eur. Phys. J. D}\ }\textbf {\bibinfo {volume}
  {66}},\ \bibinfo {pages} {157} (\bibinfo {year} {2012})}\BibitemShut
  {NoStop}%
\bibitem [{\citenamefont {Schmiegelow}\ \emph {et~al.}(2016)\citenamefont
  {Schmiegelow}, \citenamefont {Schulz}, \citenamefont {Kaufmann},
  \citenamefont {Ruster}, \citenamefont {Poschinger},\ and\ \citenamefont
  {Schmidt-Kaler}}]{schmiegelow2015excitation}%
  \BibitemOpen
  \bibfield  {author} {\bibinfo {author} {\bibfnamefont {C.~T.}\ \bibnamefont
  {Schmiegelow}}, \bibinfo {author} {\bibfnamefont {J.}~\bibnamefont {Schulz}},
  \bibinfo {author} {\bibfnamefont {H.}~\bibnamefont {Kaufmann}}, \bibinfo
  {author} {\bibfnamefont {T.}~\bibnamefont {Ruster}}, \bibinfo {author}
  {\bibfnamefont {U.~G.}\ \bibnamefont {Poschinger}}, \ and\ \bibinfo {author}
  {\bibfnamefont {F.}~\bibnamefont {Schmidt-Kaler}},\ }\bibfield  {title}
  {\bibinfo {title} {Transfer of optical orbital angular momentum to a bound
  electron},\ }\href@noop {} {\bibfield  {journal} {\bibinfo  {journal} {Nat.
  Comm.}\ }\textbf {\bibinfo {volume} {7}},\ \bibinfo {pages} {12998} (\bibinfo
  {year} {2016})}\BibitemShut {NoStop}%
\bibitem [{\citenamefont {Seymour}\ and\ \citenamefont
  {Alfano}(1980)}]{seymour1980time}%
  \BibitemOpen
  \bibfield  {author} {\bibinfo {author} {\bibfnamefont {R.}~\bibnamefont
  {Seymour}}\ and\ \bibinfo {author} {\bibfnamefont {R.}~\bibnamefont
  {Alfano}},\ }\bibfield  {title} {\bibinfo {title} {Time-resolved measurement
  of the electron-spin relaxation kinetics in {GaAs}},\ }\href@noop {}
  {\bibfield  {journal} {\bibinfo  {journal} {Appl. Phys. Lett.}\ }\textbf
  {\bibinfo {volume} {37}},\ \bibinfo {pages} {231} (\bibinfo {year}
  {1980})}\BibitemShut {NoStop}%
\bibitem [{\citenamefont {Kittel}\ and\ \citenamefont
  {Fong}(1963)}]{kittel1963quantum}%
  \BibitemOpen
  \bibfield  {author} {\bibinfo {author} {\bibfnamefont {C.}~\bibnamefont
  {Kittel}}\ and\ \bibinfo {author} {\bibfnamefont {C.-y.}\ \bibnamefont
  {Fong}},\ }\href@noop {} {\bibinfo {title} {Quantum theory of solids}},\
  Vol.~\bibinfo {volume} {3}\ (\bibinfo  {publisher} {Wiley New York},\
  \bibinfo {year} {1963})\BibitemShut {NoStop}%
\bibitem [{\citenamefont {Ridley}(2013)}]{ridley2013quantum}%
  \BibitemOpen
  \bibfield  {author} {\bibinfo {author} {\bibfnamefont {B.~K.}\ \bibnamefont
  {Ridley}},\ }\href@noop {} {\bibinfo {title} {Quantum processes in
  semiconductors}}\ (\bibinfo  {publisher} {Oxford University Press},\ \bibinfo
  {year} {2013})\BibitemShut {NoStop}%
\bibitem [{\citenamefont {Naka}\ \emph {et~al.}(2016)\citenamefont {Naka},
  \citenamefont {Morimoto},\ and\ \citenamefont {Akimoto}}]{naka2016excitons}%
  \BibitemOpen
  \bibfield  {author} {\bibinfo {author} {\bibfnamefont {N.}~\bibnamefont
  {Naka}}, \bibinfo {author} {\bibfnamefont {H.}~\bibnamefont {Morimoto}}, \
  and\ \bibinfo {author} {\bibfnamefont {I.}~\bibnamefont {Akimoto}},\
  }\bibfield  {title} {\bibinfo {title} {Excitons and fundamental transport
  properties of diamond under photo-injection},\ }\href@noop {} {\bibfield
  {journal} {\bibinfo  {journal} {Phys. Status Solidi A}\ }\textbf {\bibinfo
  {volume} {213}},\ \bibinfo {pages} {2551} (\bibinfo {year}
  {2016})}\BibitemShut {NoStop}%
\bibitem [{\citenamefont {Afanasev}\ \emph
  {et~al.}(2013{\natexlab{b}})\citenamefont {Afanasev}, \citenamefont
  {Carlson},\ and\ \citenamefont {Mukherjee}}]{Afanasev:2013kaa}%
  \BibitemOpen
  \bibfield  {author} {\bibinfo {author} {\bibfnamefont {A.}~\bibnamefont
  {Afanasev}}, \bibinfo {author} {\bibfnamefont {C.~E.}\ \bibnamefont
  {Carlson}}, \ and\ \bibinfo {author} {\bibfnamefont {A.}~\bibnamefont
  {Mukherjee}},\ }\bibfield  {title} {\bibinfo {title} {Off-axis excitation of
  hydrogenlike atoms by twisted photons},\ }\href {\doibase
  10.1103/PhysRevA.88.033841} {\bibfield  {journal} {\bibinfo  {journal} {Phys.
  Rev. A}\ }\textbf {\bibinfo {volume} {88}},\ \bibinfo {pages} {033841}
  (\bibinfo {year} {2013}{\natexlab{b}})}\BibitemShut {NoStop}%
\bibitem [{\citenamefont {Scholz-Marggraf}\ \emph {et~al.}(2014)\citenamefont
  {Scholz-Marggraf}, \citenamefont {Fritzsche}, \citenamefont {Serbo},
  \citenamefont {Afanasev},\ and\ \citenamefont {Surzhykov}}]{Scholz2014}%
  \BibitemOpen
  \bibfield  {author} {\bibinfo {author} {\bibfnamefont {H.~M.}\ \bibnamefont
  {Scholz-Marggraf}}, \bibinfo {author} {\bibfnamefont {S.}~\bibnamefont
  {Fritzsche}}, \bibinfo {author} {\bibfnamefont {V.~G.}\ \bibnamefont
  {Serbo}}, \bibinfo {author} {\bibfnamefont {A.}~\bibnamefont {Afanasev}}, \
  and\ \bibinfo {author} {\bibfnamefont {A.}~\bibnamefont {Surzhykov}},\
  }\bibfield  {title} {\bibinfo {title} {Absorption of twisted light by
  hydrogenlike atoms},\ }\href {\doibase 10.1103/PhysRevA.90.013425} {\bibfield
   {journal} {\bibinfo  {journal} {Phys. Rev. A}\ }\textbf {\bibinfo {volume}
  {90}},\ \bibinfo {pages} {013425} (\bibinfo {year} {2014})}\BibitemShut
  {NoStop}%
\bibitem [{\citenamefont {Afanasev}\ \emph {et~al.}(2016)\citenamefont
  {Afanasev}, \citenamefont {Carlson},\ and\ \citenamefont
  {Mukherjee}}]{afanasev2016high}%
  \BibitemOpen
  \bibfield  {author} {\bibinfo {author} {\bibfnamefont {A.}~\bibnamefont
  {Afanasev}}, \bibinfo {author} {\bibfnamefont {C.~E.}\ \bibnamefont
  {Carlson}}, \ and\ \bibinfo {author} {\bibfnamefont {A.}~\bibnamefont
  {Mukherjee}},\ }\bibfield  {title} {\bibinfo {title} {High-multipole
  excitations of hydrogen-like atoms by twisted photons near a phase
  singularity},\ }\href@noop {} {\bibfield  {journal} {\bibinfo  {journal} {J.
  Opt.}\ }\textbf {\bibinfo {volume} {18}},\ \bibinfo {pages} {074013}
  (\bibinfo {year} {2016})}\BibitemShut {NoStop}%
\bibitem [{\citenamefont {Dunham}(1993)}]{dunham1993investigations}%
  \BibitemOpen
  \bibfield  {author} {\bibinfo {author} {\bibfnamefont {B.~M.}\ \bibnamefont
  {Dunham}},\ }\bibinfo {title} {Investigations of the physical properties of
  photoemission polarized electron sources for accelerator applications},\
  \href@noop {} {Ph.D. thesis},\ \bibinfo  {school} {University of Illinois at
  Urbana-Champaign} (\bibinfo {year} {1993})\BibitemShut {NoStop}%
\bibitem [{\citenamefont {Afanasev}\ \emph
  {et~al.}(2018{\natexlab{b}})\citenamefont {Afanasev}, \citenamefont
  {Carlson},\ and\ \citenamefont {Solyanik}}]{afanasev2018E2M1}%
  \BibitemOpen
  \bibfield  {author} {\bibinfo {author} {\bibfnamefont {A.}~\bibnamefont
  {Afanasev}}, \bibinfo {author} {\bibfnamefont {C.~E.}\ \bibnamefont
  {Carlson}}, \ and\ \bibinfo {author} {\bibfnamefont {M.}~\bibnamefont
  {Solyanik}},\ }\bibfield  {title} {\bibinfo {title} {Atomic spectroscopy with
  twisted photons: Separation of {$M1\ensuremath{-}E2$} mixed multipoles},\
  }\href {\doibase 10.1103/PhysRevA.97.023422} {\bibfield  {journal} {\bibinfo
  {journal} {Phys. Rev. A}\ }\textbf {\bibinfo {volume} {97}},\ \bibinfo
  {pages} {023422} (\bibinfo {year} {2018}{\natexlab{b}})}\BibitemShut
  {NoStop}%
\bibitem [{\citenamefont {Syouji}\ \emph {et~al.}(2017)\citenamefont {Syouji},
  \citenamefont {Saito},\ and\ \citenamefont {Otomo}}]{syouji2017creation}%
  \BibitemOpen
  \bibfield  {author} {\bibinfo {author} {\bibfnamefont {A.}~\bibnamefont
  {Syouji}}, \bibinfo {author} {\bibfnamefont {S.}~\bibnamefont {Saito}}, \
  and\ \bibinfo {author} {\bibfnamefont {A.}~\bibnamefont {Otomo}},\ }\bibfield
   {title} {\bibinfo {title} {Creation of excitons excited by light with a
  spatial mode},\ }\href@noop {} {\bibfield  {journal} {\bibinfo  {journal} {J.
  Phys. Soc. Jpn.}\ }\textbf {\bibinfo {volume} {86}},\ \bibinfo {pages}
  {124720} (\bibinfo {year} {2017})}\BibitemShut {NoStop}%
\end{thebibliography}%
\bibliographystyle{apsrev4-1_loc}

\end{document}